\documentclass{my-alternate-2013}

\usepackage{url}
\usepackage{color}
\usepackage{algorithm}
\usepackage{multirow}
\usepackage[rm]{caption}
\DeclareCaptionType{copyrightbox}
\usepackage{threeparttable}
\usepackage{booktabs}
\usepackage{xspace}
\usepackage{algpseudocode}
\usepackage{times}
\usepackage{balance}

\usepackage[square,comma,numbers,sort]{natbib}

\newcommand{\metricfont}[1]{{\small\sf{#1}}}
\newcommand{\metric}[1]{\metricfont{#1}}
\newcommand{\MED}{\metric{MED}}

\newcommand{\med}[1]{{\ensuremath\metric{MED}_{\small\metricfont{#1}}}}

\def\D{\hphantom{1}}
\def\C{\hphantom{1,}}
\newcommand\method[1]{{\sf\small{#1}}}

\newcommand{\tfidf}{\mbox{\textsc{tf}$\times$\textsc{idf}}\xspace}

\newcommand{\okapi}{\mbox{\scriptsize BM25}\xspace}
\newcommand{\lm}{\mbox{\scriptsize LM}\xspace}

\newcommand{\tfbm}{\ensuremath{\mbox{TF}_{\mbox{\scriptsize{BM25}}}\xspace}}

\newcommand{\indri}{\method{Indri\xspace}}

\newcommand{\var}[1]{\mbox{\emph{#1}}}
\def\D{\hphantom{1}}
\def\C{\hphantom{1,}}
\newcommand{\myurl}[1]{{\url{#1}}}

\newcommand{\myparagraph}[1]{\vspace{0.6\baselineskip}\noindent{\textbf{#1}}~}

\newcommand{\mycomment}[1]{}
\newcommand{\mylabel}[1]{\label{#1}}

\newcommand{\cmath}{\ensuremath{\mathcal{C}}}

\newlength{\onedigit}
\algrenewcommand{\algorithmiccomment}[1]{\hfill// #1}
\algnewcommand{\LineComment}[1]{\State// #1}
\algrenewcommand{\algorithmicrequire}{\textbf{Input:}}
\algrenewcommand{\algorithmicensure}{\textbf{Output:}}

\newcommand{\onecolfig}{.35}
\newcommand{\onecolfigl}{.5}

\begin{document}


\title{Dynamic Trade-Off Prediction in Multi-Stage
Retrieval Systems}






\clubpenalty=10000
\widowpenalty = 10000


\numberofauthors{3} \author{
\alignauthor J. Shane Culpepper\\[0.5ex]
\affaddr{RMIT University\\
Melbourne, Australia\\[0.5ex]
shane.culpepper@rmit.edu.au}
\and
\alignauthor Charles L. A. Clarke\\[0.5ex]
\affaddr{University of Waterloo\\
Waterloo, Canada\\[0.5ex]
claclarke@cs.uwaterloo.ca}
\and
\alignauthor Jimmy Lin\\[0.5ex]
\affaddr{University of Waterloo\\
Waterloo, Canada\\[0.5ex]
jimmylin@uwaterloo.ca}
}

\maketitle

\begin{abstract}

Modern multi-stage retrieval systems are comprised of a candidate
generation stage followed by one or more reranking stages.
In such an architecture, the quality of the final ranked list may not
be sensitive to the quality of initial candidate pool, especially in
terms of early precision.
This provides several opportunities to increase retrieval efficiency
without significantly sacrificing effectiveness.
In this paper, we explore a new approach to dynamically predicting
two different parameters in the candidate generation stage which can
directly affect the overall efficiency and effectiveness of the
entire system.
Previous work exploring this tradeoff has focused on global
parameter settings that apply to all queries, even though optimal
settings vary across queries.
In contrast, we propose a technique which makes a parameter
prediction that maximizes efficiency within a effectiveness envelope
on a {\it per query} basis, using only static pre-retrieval features.
The query-specific tradeoff point between effectiveness and
efficiency is decided using a classifier cascade that weighs
possible efficiency gains against effectiveness losses over a range
of possible parameter cutoffs to make the prediction.
The interesting twist in our new approach is to train classifiers
without requiring explicit relevance judgments.
We show that our framework is generalizable by applying it to two
different retrieval parameters --- selecting $k$ in common top-$k$ 
query retrieval algorithms, and setting a quality threshold, $\rho$,
for score-at-a-time approximate query evaluation algorithms.
Experimental results show that substantial efficiency gains are
achievable depending on the dynamic parameter choice.
In addition, our framework provides a versatile tool that can be used
to estimate the effectiveness-efficiency tradeoffs that are possible
{\em before} selecting and tuning algorithms to make machine learned
predictions.

\end{abstract}





\section{Introduction}

Effectiveness-efficiency tradeoffs have been extensively explored
in search engine architectures:\ Highly-effective ranking models
often take advantage of computationally expensive features and hence
are slow, while fast ranking algorithms often sacrifice effectiveness.
In a modern multi-stage ranking architecture~\citep{p10-query,czc10-wsdm,mso13irj,msoh13acmtois,al13irj,al13sigir},
an initial candidate generation stage is followed by one or more
rerankers, and the end-to-end effectiveness-efficiency tradeoffs are
often a combination of many different component-level tradeoffs.
In this work, we focus on the initial candidate generation stage,
whose responsibility is to provide an initial set of documents that
are then considered in more detail, for example, by
machine-learned rankers~\citep{b05-icml,b10-learning}.
Previous work~\cite{Asadi_Lin_CIKM2012} has shown that the quality of the final ranked
list is relatively insensitive to the quality of the initial
candidate set, especially in terms of early precision.
The intuition is as follows:\ as long as the candidate generation stage can supply
a ``reasonable'' pool of documents, it is likely that later-stage
rankers can identify the best documents and place them in high
ranking positions, regardless of the original rank scores.
If the final ranked list is assessed in terms of, say, NDCG@10,
the initial candidate pool only needs to contain ten documents of the
highest relevance grade to achieve the best possible score---provided
that the later-stage rankers identify these documents (which is
likely, given the sophistication of modern machine-learned rankers).

The relative lack of sensitivity to the initial candidate pool
creates opportunities to increase efficiency without sacrificing
effectiveness in the candidate generation stage---in the sense that
we can ``cut corners'' without impacting the quality of the final
ranked list.
This is the focus of our work.
We explore two orthogonal approaches to tuning the
effectiveness-efficiency tradeoff.
The first is the size of the candidate pool $k$.
In a standard document-at-a-time query evaluation algorithm, query
evaluation latency increases as a function of $k$.
A large candidate document pool also means greater cost in the
feature extraction and reranking stages downstream.
Thus, we desire a $k$ as small as possible while remaining within an
effectiveness envelope.
The second approach is to take advantage of score-at-a-time {\it
approximate} query evaluation strategies.
In particular, we adopt a publicly available technique that comes with
a ``quality knob'' called
$\rho$.\footnote{\url{https://github.com/lintool/JASS}}
For any fixed cutoff, we can control the retrieval quality (with respect
to exhaustive query evaluation) by adjusting $\rho$.
An obvious third step would be to tune both $k$ and quality at the
same time, but we leave this for future work since it significantly
increases the decision space of the classifier.

\myparagraph{Our Contributions.}  The key contribution of our work is
to show that we can achieve substantial savings in candidate
generation efficiency in multi-stage ranking {\it without sacrificing effectiveness}, tuned on a 
{\it per query} basis, using only static pre-retrieval features {\it without
requiring relevance judgments}.
We accomplish this by building classifier cascades that make binary
decisions at several different cutoffs along an effectiveness-efficiency
tradeoff curve (using either parameters $k$ or $\rho$).
In effect, each classifier in the cascade weighs possible efficiency
gains against effectiveness losses and either decides to ``take
action'' (by selecting the current parameter cutoff) or pass the
decision to the next stage in the cascade.

It is true that several other recent works have investigated tradeoffs between effectiveness
and efficiency in multi-stage ranking architectures
~{\citep{czc10-wsdm,Asadi_Lin_CIKM2012,al13irj,al13sigir,mso13irj,msoh13acmtois}}.
However, previous work has mostly focused on finding global settings across
a collection of queries, and do not focus on query-sensitive cutoffs as we do here.
In addition, a key feature of our approach, worth emphasizing, is that we are able to train these classifier cascades {\it
without requiring relevance judgments}, which overcomes a limitation with almost
all previous studies since relevance judgments restrict the scope of their experiments to at most
a few hundred queries. In contrast, we are able to run experiments on tens of thousands of queries.
This can be achieved by leveraging a recently introduced
evaluation technique called Maximized Effectiveness Difference (\MED)~{\citep{tc15,ccm16irj}}.
We apply our general classifier cascade framework to two completely
different query evaluation algorithms:\ tuning $k$ in a standard
document-at-a-time {\textsc{Wand}} algorithm, and tuning the quality
parameter $\rho$ in a recently developed score-at-a-time approximation
algorithm.
Our experimental results show a $50\%$ or more improvement in
efficiency without any significant loss in effectiveness.
The fact that our framework generalizes to two different 
approaches to candidate generation in multi-stage ranking
highlights its flexibility and generality.

\section{Background}
\label{sec-background}

We assume a standard formulation of the ranked retrieval problem,
where given a user query $q$, our goal is to return a ranked list
that maximizes a particular metric. In the web context, the metric
would likely emphasize early precision, e.g., NDCG@10. In this section,
we discuss tradeoffs between effectiveness and efficiency in the
context of multi-stage ranking.

\subsection{Multi-Stage Ranking Efficiency}
\label{sec21-msr}

\begin{figure}[t]
\centering
\includegraphics[width=0.45\textwidth]{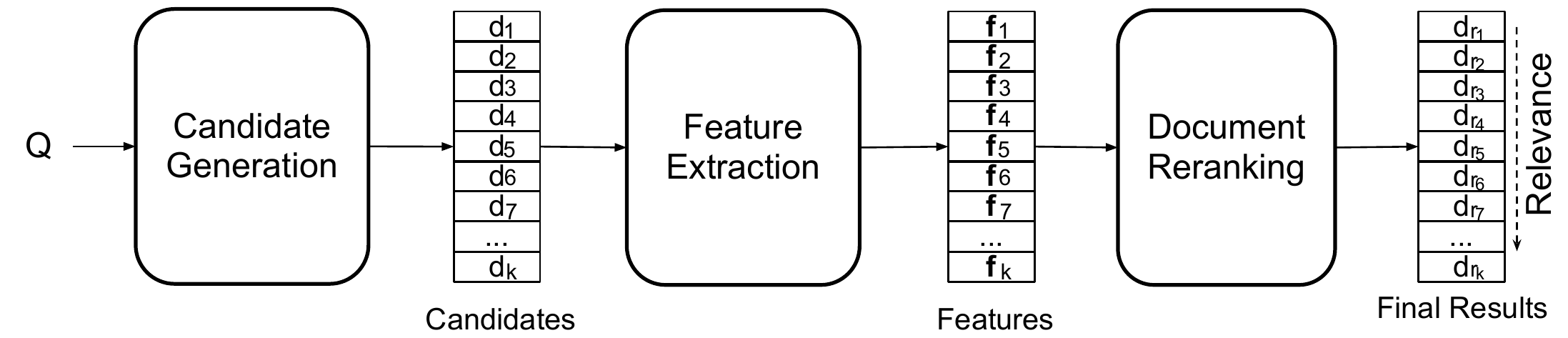}
\caption{Illustration of a multi-stage retrieval architecture with
  distinct candidate generation, feature extraction, and document
  reranking stages.
\label{figure:ltr_architecture}}
\end{figure}

Multi-stage retrieval systems have become the dominant
model for efficient and effective web search
~{\citep{p10-query,czc10-wsdm,mso13irj,msoh13acmtois,al13irj,al13sigir}}.
The key idea of this approach is to efficiently generate a set of
candidate documents that are likely to be relevant to a query, and
then iteratively reorder the documents using a series of more
expensive machine learning techniques.
As the cost of the later stage reordering can be computationally
expensive, minimizing the number of candidate documents in early
stage retrieval can yield significant benefits in overall query
processing time.
{\citet{kdf13-kdd}} showed that every $100$ ms boost in search speed
increases revenue by $0.6\%$ at Bing.
So, even small gains in overall performance can translate to tangible
benefits in commercial search engines.

\begin{figure}[t]
\centering

\includegraphics[width=\onecolfig\textwidth]{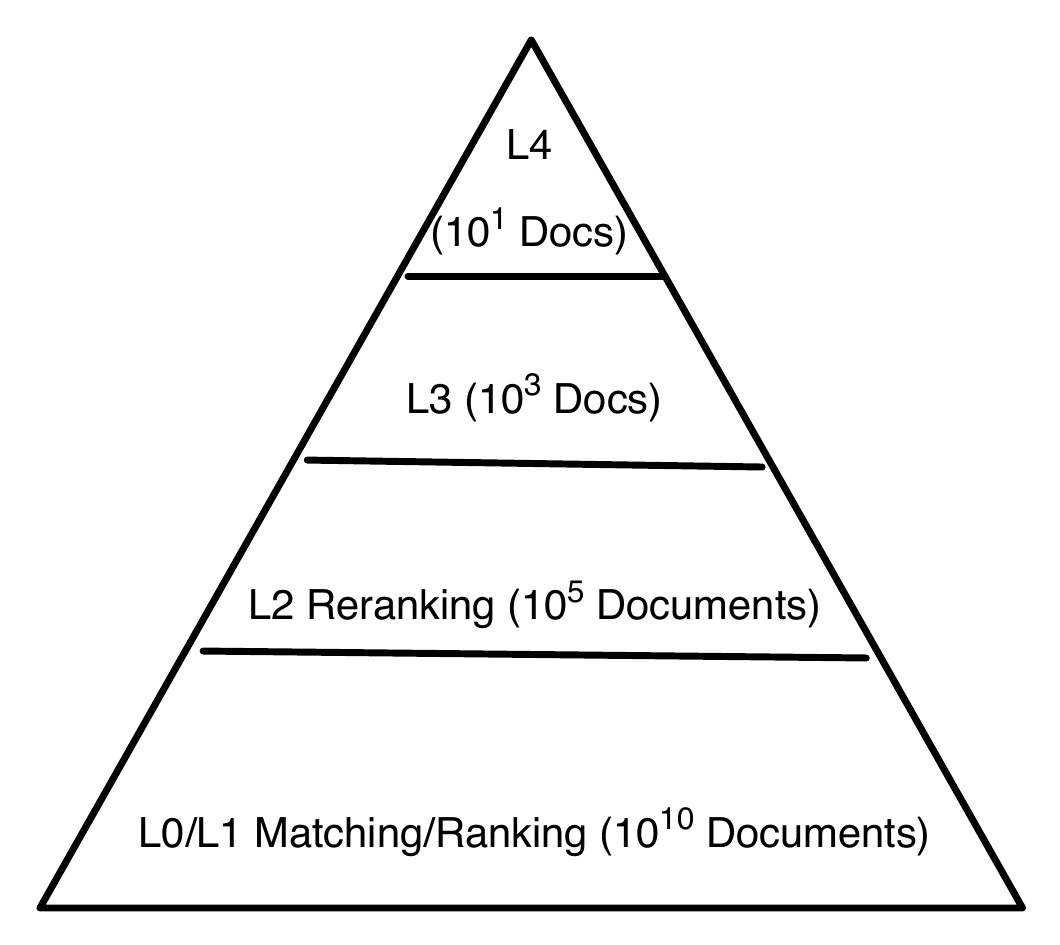}
  \caption{A four stage retrieval process as originally described
  by {\citet{p10-query}}.
  }
  \label{fig:mstage}
\end{figure}

The simplest example of a multi-stage ranking architecture is
illustrated in Figure~\ref{figure:ltr_architecture}.
The input to the candidate generation stage is a query $q$ and the
output is a set of $k$ document ids $\{d_1, d_2, \ldots d_k\}$.
In principle, the candidate pool can be treated as a ranked list or a
set---the difference is whether subsequent stages take advantage of
the document score or ranking.
These document ids serve as input to the feature extraction stage,
which returns a list of $k$ feature vectors $\{\mathbf{f}_1,
\mathbf{f}_2, ... \mathbf{f}_k\}$, each corresponding to a candidate
document. These serve as input to the document reranking stage,
which typically applies a machine-learned model to produce a final
ranking. Of course, there can be an arbitrary number of reranking
stages.
For example, \citet{p10-query} describes a four-stage retrieval
architecture in Bing, as shown in Figure~{\ref{fig:mstage}}.
The key take-away message is that increasingly expensive reranking
steps benefit from processing fewer and fewer documents.

It is important to emphasize that the size of the candidate pool of
documents $k$ is completely independent of the size of the final
ranked list (with only the hard constraint that the final size has to
be at least $k$).
So, what is the proper setting of $k$?
The work of {\citet{mso13irj}} suggest several potential answers ---
``tens of thousands'' ({\citet{cc11-jmlr}}), $5{,}000$
({\citet{cf+09-trec}}), $1{,}000$ ({\citet{ql+09-ir}}, or smaller
samples such as $200$ ({\citet{zk+09-lr4ir}}) or even $20$
({\citet{czc10-wsdm}}).
Of course, the larger the $k$, the slower the system, in two
respects:
First, in standard document-at-a-time query evaluation algorithms
that would provide the initial candidate documents (e.g.,
\textsc{Wand}), $k$ has a direct impact on query latency, since
a larger heap needs to be maintained, providing fewer opportunities
for early exits depending on document score distributions.
Second, for every document in the candidate pool, we need to run the
feature extractors to serve as input to the subsequent reranking
stages (see Figure~\ref{figure:ltr_architecture}).
Thus, from an efficiency perspective, it is clear that we desire the
smallest possible $k$ that allows end-to-end effectiveness to remain
within some bounded envelope (see below for more details).

Note that our work focuses on the size of the candidate pool for the
purposes of ranking at query time.
In contrast, {\citet{mso13irj}} focused on the importance of
candidate pool size for the {\it training} of learning-to-rank
systems.
In particular, they looked at how the size of the candidate set
effects the final results, arguing that the relationship is dependent
on the type of information need.
They show that as few as $10$-$20$ documents may be needed for TREC
2009 and 2010 web track queries, but as many as $1{,}500$ may be
needed for navigational information needs on the same corpus.
They also argue that field features such as anchor text are critical
in the first stage retrieval process.
Since we are concerned with the application of machine-learned models
at runtime, the work of \citet{mso13irj} is somewhat orthogonal
to our study.

In terms of candidate generation, being able to control $k$ can have
a substantial impact on end-to-end efficiency.
An alternative approach might be, for a fixed $k$, to take advantage
of {\it approximate} query evaluation algorithms that trade off the
quality of the retrieved results for efficiency.
As an example, \citet{Asadi_Lin_CIKM2012} devised posting list
intersection algorithms that take advantage of Bloom filters to
generate result sets very quickly, but suffer from false positives,
i.e., a retrieved document may not actually have all the query terms.

In this work, we explore a recently-developed and publicly available
score-at-a-time approximate query evaluation algorithm proposed by
{\citet{Lin_Trotman_ICTIR2015}} referred to as JASS.
Instead of computing floating point document scores, their technique
used quantized impact scores~\cite{Anh_etal_SIGIR2001}, which
increases query evaluation speed by replacing floating point
operations with simple integer additions to accumulate document
scores while traversing postings.
The query evaluation algorithm takes advantage of impact-ordered
indexes to process posting segments in decreasing score order.
Since contributions to the document scores monotonically decrease,
query evaluation can quit at any time.
Early termination is controlled by a parameter called $\rho$, which
is simply the number of postings to be processed.
As $\rho$ increases, the ranked list approaches that of exhaustive
evaluation, which produces a precise ranking based on document
scores.

{\citet{Lin_Trotman_ICTIR2015} show that $\rho$ correlates linearly
with wall-clock query evaluation time, with an $R^2$ value of over
$0.9$ on Clue-Web09 and ClueWeb12 data.
They further suggest that a $\rho$ setting equal to $10\%$ of the size of
the collection achieves the best compromise between effectiveness and
efficiency.
Indeed, in a recent large-scale evaluation of open-source search
engines~\cite{Arguello_etal_FORUM}, this new score-at-a-time
approach was shown to be the fastest among all submissions.
The suggested setting of $\rho$, however, was not fully explored, and
furthermore is currently a global setting used for all queries.

It is clear that for candidate generation in multi-stage ranking, $k$
in \textsc{Wand} and $\rho$ in JASS represent important efficiency
``knobs''.
We would like to set both values as low as possible, but not 
sacrifice end-to-end effectiveness.
This is, in short, the story and goal of this paper.
We show that it is possible to predict, on a per-query basis, a
minimum $k$ and $\rho$ such that end-to-end effectiveness remains
within a bounded envelope, purely based on pre-retrieval static
features, without requiring any relevance judgments.

The closest related work to ours is that of
{\citet{tmo13-wsdm}, who also attempt to tune effectiveness-efficiency
tradeoffs on a per query basis using query difficulty and query
efficiency prediction techniques. However, their choice of settings
is rather coarse grained:\ they only select between two
configurations, whereas our classifier cascades are able to consider
many more settings. Furthermore, their work exhibits the same limitation
as most previous studies in requiring relevance judgments for training,
and hence they are only able to experiment on 150 queries
from TREC 2009--2011. In contrast, since our approach does not require
any relevance judgments, we can tune our techniques on tens of thousands
of queries, as we will discuss next.

\subsection{Multi-Stage Ranking Effectiveness}
\label{sec22-effectiveness}

\begin{figure}[t]
\centering
\includegraphics[width=\onecolfigl\textwidth]{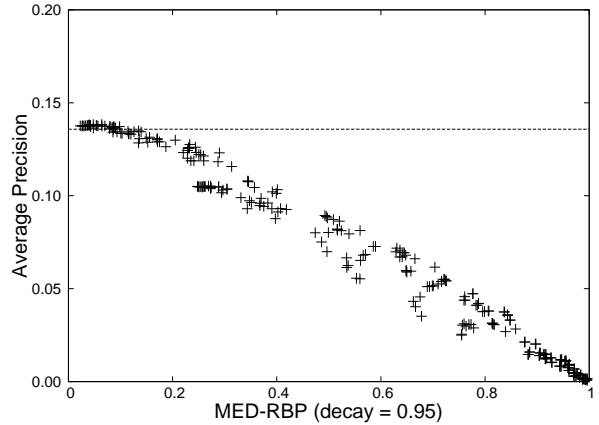}
  \caption{
Correlation between $\med{RBP}$ and measured average precision using the
50~queries of the TREC 2010 Web Track adhoc collection and
the {\tt{IvoryL2Rb}} experimental run as the second-stage.
Each of the $279$~points represents one of $31$ distinct first stages,
across a range of parameter settings.
The dashed line indicates the performance of the second stage with no
early-stage filter.
Reproduced from {\citet{tc15}}.
  }
  \label{fig:med}
\end{figure}

Tuning ranking parameters requires substantial training data to measure
effectiveness.
Fortunately, in the case of tuning candidate generation for a second-stage
ranker (see Figure~\ref{figure:ltr_architecture}), we have this training data readily available,
since the second stage itself may be enlisted to provide it.
To create this training data we first run the second stage ranker
over a very large candidate set, much larger than time might allow
for interactive search.
Conceptually this candidate set might be the entire collection, but
practically it will be limited to a subset retrieved from query
keyword matches and other simple features.
Ideally, this set would contain all relevant documents, but mixed
together with many non-relevant documents.

The second-stage ranker then ranks this set, producing a ranked list $A$.
Given the potential size of the set, producing this ranking may take
substantial time.
However, while this time may be far greater than would be tolerable
for interactive searching, when creating training data, time is not a
problem.

Now, suppose we have a more efficient candidate generation algorithm,
designed to feed this second stage ranker.
It produces a much smaller set, which can be more efficiently ranked
by the second stage to produce a ranked list $B$.
We measure the effectiveness of the candidate generation algorithm
according to its ability to {\em supply the documents that the second
stage needs} in the absence of efficiency constraints, i.e., $A$.
More specifically, we compute a rank correlation coefficient or rank
similarity measure between $A$ and $B$, using its value $S(A,B)$ as
our effectiveness metric.

Naturally, the similarity measure must be suitable for this
purpose~\cite{wmz10acmtois}.
In particular, a rank similarity measure for search results must be
appropriately {\em top-weighed}, placing greater emphasis on earlier
ranks than on later ranks.
If the top document in $A$ is missing from $B$, the impact on the
user will be much greater than if the $100$th document is missing.

{\citet{tc15}} describe a family of rank similarity measures
specifically intended for comparing ranked lists produced by search
engines.
Given a traditional effectiveness metric~---~such as MAP~\cite{gmap},
RBP~\cite{mz08acmtois}, DCG~\cite{jk02acmtois}, or
ERR~\cite{cmzg09cikm}~---~ {\citeauthor{tc15}} define a distance
measure between two ranked lists in terms of that metric, as follows:
``Given two ranked lists, $A$ and $B$, what is the maximum difference
in their effectiveness scores possible under [that metric].''

They call this family of distance measures {\em maximized
effectiveness difference} (\mbox{MED$(A,B)$}) and develop variants
corresponding to several standard effectiveness metrics~---~including
$\med{MAP}$, $\med{RBP}$, $\med{DCG}$, and $\med{ERR}$.
They explore {\MED} as a method for quantifying changes to ranking
algorithms without the need for human relevance judgments.
For example, {\MED} allows a search to identify queries for which
a proposed change causes the greatest impact.
An open-source implementation is publicly available online,{\footnote{\url{https://github.com/claclark/MED}}}
which can be used to compute {\MED} for various effectiveness
measures, and is used in this paper.

Building on this work, we have recently applied {\MED} to measure
effectiveness of the initial stages in multi-stage rankers~{\citep{ccm16irj}}.
That work follows the procedure outlined above, using a second-stage
ranking as a gold standard to measure first-stage effectiveness,
validating this procedure.
Unlike previous explorations of efficiency-effectiveness tradeoffs,
the absence of any requirement for human relevance judgments allows
the procedure to be easily applied across tens of thousands of queries.

For illustration purposes, Figure~\ref{fig:med} is reproduced from that paper.
The figure shows the performance of a number of first-stage rankers,
operating over a range of parameter settings,
supplying a high-quality second-stage ranker.
The horizonal dashed line indicates the performance of the second-stage ranker
without first-stage filtering.
Values of $\med{RBP}$ below 0.05 produce no practical loss in
measured effectiveness.

In this range, {\MED} is measuring shortcomings in the first-stage that are
not necessarily reflected in the evaluation measures. 
A first-stage ranker that fails to return the top document required by the
second stage will receive a lower {\MED} score than a first-stage ranker
that fails to return the sixth document.
While other relevant documents might move up to replace the lost
documents, leaving both with the same measured effectiveness,
losing the top-ranked document is viewed more seriously than losing
lower-ranked documents.

Previous work explored efficiency-effectiveness tradeoffs of
first-stage algorithms, and their parameter settings, as applied
uniformly across all queries (including the blinded citation above).
However, optimal algorithms and settings vary across queries.
In this paper we explore a technique for optimizing
efficiency-effectiveness tradeoffs on a per-query basis, selecting
the optimal algorithm and setting for each using static,
pre-retrieval features.
While we focus specifically on the two-stage architecture in
Figure~\ref{figure:ltr_architecture}, our methods should generalize
to larger multistage architectures,
such as shown in Figure~{\ref{fig:mstage}},
with the effectiveness of each stage measured in terms of the next.

\section{Approach}
\label{sec-approach}

\myparagraph{Feature Selection.}
Simple term features have been used successfully in a variety
of different learning to rank scenarios
~{\citep{l09-ltr,mso12-cikm,mso13irj,msoh13acmtois}} and in
query difficulty prediction~{\citep{cyt10-qpb,mto12-sigir,kh+15-wsdm}}.
Across all of this work one general theme has emerged -- a mixture
of similarity measures and query specific score aggregation techniques
yield the most benefit.
Inspired by this previous work, we adopt this philosophy in our
feature choices as well.
We use three simple similarity measures in this work:
\begin{enumerate}
\item BM25 with the formulation:

\[
  \D{\okapi} = \log \left( \frac{N - f_t + 0.5}{f_t + 0.5} \right) \cdot {\tfbm}
\]

\[
  {\tfbm} = \frac{f_{t,d} \cdot (k_1 + 1)}{f_{t,d} + k_1 \cdot ((1 - b) +
  (b \cdot \ell_d/\ell_{avg}))}
\]

where $N$ is the number of documents in the collection, $f_t$ is the
number of distinct document appearances of $t$, $f_{d,t}$ is the number of
occurrences of term $t$ in document $d$, $k_1 = 0.9$, $b = 0.4$
\footnote{The values for $b$ and $k_1$ are different
than the defaults reported by {\citet{rwj+94-trec}}.
These parameter choices were reported for Atire and Lucene
in the 2015 IR-Reproducibility Challenge, see
{\url{github.com/lintool/IR-Reproducibility}} for further
details.},
$\ell_d$ is the number of terms in document $d$, and
$\ell_{avg}$ is the average of $\ell_d$ over the whole collection.

\item Query Likelihood using a Dirichlet prior smoothing
formulation:
\[
  \D{\lm} = \log \left( \frac{f_{t,d} + \mu \cdot C_t / |C|}{\ell_d + \mu} \right)
\]

where $C_t$ is the number of occurrences of $t$ in the collection,
$|C|$ is the length of the collection (total number of terms), $\mu =
2500$ is the smoothing factor, and all other variables are the same
as in {\okapi}.

\item {\tfidf} with the formulation:
\[
  \D{\tfidf} = \frac{1}{\ell_d} \cdot (1 + \log (f_{t,d})) \cdot \log (1 + \frac{N}{f_t})
\]

\end{enumerate}

\begin{table}[t]
\centering
\begin{tabular}{l}
\toprule
\multicolumn{1}{c}{Term Statistics} \\
\midrule
1.  Number of occurrences of term $t$ in collection ($C_t$). \\
2.  Number of documents containing term $t$ ($f_t$). \\
3.  Maximum Similarity Score \\
4.  First Quartile Similarity Score \\
5.  Third Quartile Similarity Score \\
6.  Minimum Similarity Score \\
7.  Arithmetic Mean of Similarity Scores \\
8.  Harmonic Mean of Similarity Scores \\
9.  Median of Similarity Scores \\
10. Variance of Similarity Scores \\
11. Interquartile Range of Similarity Scores \\
\bottomrule
\end{tabular}
\caption{Query independent term features used by the classifier.
Each feature can be precomputed and stored with the postings list.
\label{tbl:tfeatures}
}
\end{table}

\begin{table}[t]
\centering
\begin{tabular}{p{72mm}}
\toprule
\multicolumn{1}{c}{Query Features (Score Dependent)} \\
\midrule
1. Arithmetic Mean of $t_f$ \\
2. Harmonic Mean of Maximum Scores \\
3. Arithmetic Mean of Maximum Scores \\
4. Arithmetic Mean of Median Score \\
5. Arithmetic Mean of Mean Scores \\
6. Arithmetic Mean of Score Variances \\
7. Arithmetic Mean of Score Interquartile Ranges \\
8. Minimum Score of terms in the query for each feature in 
   Table~{\ref{tbl:tfeatures}}. \\
9. Maximum Score of terms in the query for each feature in 
   Table~{\ref{tbl:tfeatures}}. \\
\midrule
\multicolumn{1}{c}{Query Features (Score Independent)} \\
\midrule
1.  Query Length \\
\bottomrule
\end{tabular}
\caption{Query specific features used by the classifier.
All score dependent features can be computed on the fly
for all three similarity metrics at query time using
the prestored values in Table~{\ref{tbl:tfeatures}}.
\label{tbl:qfeatures}
}
\end{table}

\begin{table*}[t]
\centering
\begin{tabular}{cccccccccc}
\toprule
\multirow{2}{*}{Topic} & \multicolumn{9}{c}{$k$}\\
&$20$&$50$&$100$&$200$&$500$&$1{,}000$&$2{,}000$&$5{,}000$&$10{,}000$\\
\midrule
20001&0.544&0.346&0.104&0.056&0.010&0.002&0.001&0.000&0.000\\
20002&0.536&0.142&0.053&0.016&0.002&0.000&0.000&0.000&0.000\\
20003&0.865&0.856&0.810&0.773&0.706&0.684&0.582&0.122&0.000\\
20004&0.999&0.944&0.132&0.070&0.018&0.008&0.008&0.000&0.000\\
\bottomrule
\end{tabular}
\caption{The ${\med{RBP}}$ scores for the first four topics in the
TREC MQ2009 collection at $9$ different cutoffs for $k$.
\label{tbl:kexample}
}
\end{table*}

These similarity formulations were used since each can easily be
precomputed for all term---document combinations and treated as
independent term-specific features.
In addition to the three similarity scoring regimes, we also
adopt several different score aggregation techniques, and compute
a variety of static statistical features for each term posting:
maximum score, minimum score, arithmetic mean of scores, harmonic
mean of scores, median of scores, variance of scores, first quartile
score, and third quartile score.
Additional query specific features are also incorporated into the
model including query length, minimum and maximum score for the terms
in the query, and means (arithmetic and harmonic) of the query specific
term scores.
Table~{\ref{tbl:tfeatures}} provides a comprehensive breakdown of the
term specific features, each of which can be computed at index time. 
Table~{\ref{tbl:qfeatures}} shows how each of the term specific
features are combined into the final feature set used by the classifier.
A total of $70$ features are used in our work.

\myparagraph{Labeling Instances.}
We now turn our attention to how the training collection was created.
One of the key ideas of this work is to use {\MED} to determine a
minimal candidate set that also maximizes the possible effectiveness
in the final reranking stage.
In order to achieve this, we have created a gold standard set 
using $40{,}000$ queries from the 2009 TREC Million Query Track.
For each query, $\med{RBP}$, $\med{ERR}$, and $\med{DCG}$ is
computed for the $k$ values of $20$, $50$, $100$,
$200$, $500$, $1{,}000$, $2{,}000$, $5{,}000$, and $10{,}000$.
Our gold standard run for tuning $k$ was the {\tt{uogTRMQdph40}} run, as it
represents one of the top-scoring systems (when measured over the small
subset of the queries that were evaluated) that returned results for all
$40{,}000$ of the MQ2009 queries.
For $\rho$, the cutoff values were $100$k, $200$k, $500$k, $1$m, $2$m,
$5$m, $10$m, $20$m, and $50$m. Our gold standard run for tuning $\rho$
is the ranked list that results from exhaustive query evaluation, which
generates an exact ranking.

So, in total we have computed {\MED} using three different metrics
at $9$ distinct cutoffs for $k$ and $\rho$.
To label the instances, we now select a sufficiently low value of
a given metric, say $\med{RBP} \le 0.05$, and choose the minimal
cutoff that satisfies this constraint---this is what we have previously
referred to as the ``effectiveness envelope'' we would like to maintain.

For example, consider the $\med{RBP}$ computations for the first $4$ topics
shown in Table~{\ref{tbl:kexample}}.
If the minimal acceptable score is $\med{RBP} \le 0.05$, then for Topic
$20001$, the nominal class assigned would be $k=500$, whereas Topic
$20002$ can achieve a similar score with $k=200$. 

\begin{figure}

\def\hd{\kern4pt}

\[
\cmath =
\begin{pmatrix}
\hd 0 & 1 & 1 & 1 & 1 & 1 & 1 & 1 & 1 \hd\\
\hd 2 & 0 & 1 & 1 & 1 & 1 & 1 & 1 & 1 \hd\\
\hd 3 & 2 & 0 & 1 & 1 & 1 & 1 & 1 & 1 \hd\\
\hd 4 & 3 & 2 & 0 & 1 & 1 & 1 & 1 & 1 \hd\\
\hd 5 & 4 & 3 & 2 & 0 & 1 & 1 & 1 & 1 \hd\\
\hd 6 & 5 & 4 & 3 & 2 & 0 & 1 & 1 & 1 \hd\\
\hd 7 & 6 & 5 & 4 & 3 & 2 & 0 & 1 & 1 \hd\\
\hd 8 & 7 & 6 & 5 & 4 & 3 & 2 & 0 & 1 \hd\\
\hd 9 & 8 & 7 & 6 & 5 & 4 & 3 & 2 & 0 \hd\\
\end{pmatrix}
\]

\caption{A nine class cost matrix which penalizes underpredictions
in a classifier.}
\label{fig:cmatrix}
\end{figure}

\myparagraph{Multilabel Classification and Regression.}
The most obvious solution from a machine learning perspective is to
train a multilabel classifier, or use the true cutoff values
with a regression algorithm such as a Reduced Error Pruning Tree
(\method{REPTree})~{\citep{ek01-reptree}}, 
a Multilayer Perceptron, or Sequential Minimal Optimization 
(\method{SMOReg})~{\citep{smoreg}}.
We explored all of these possibilities in our early empirical
analysis, and found that none of the approaches was reliably 
better than using a fixed cutoff baseline.

After careful examination of the initial results, a clear constraint emerged
in producing good results in our classifier --- any under-prediction
(False Positives) can significantly hurt overall effectiveness and
should be avoided.
A standard approach to reweight classification is to use
a cost sensitive classifier~{\citep{e01-ijcai}} such as
{\method{MetaCost}~{\citep{d99-kdd}}.
Our experiments with a cost sensitive classifier that penalized the
classifier for under-predicting (false positives) were promising.
For example, using the cost matrix $\cmath$ shown in
Figure~{\ref{fig:cmatrix}}} provides a better solution than either a
multilabel classification or a regression.
At the bottom of the matrix, we penalize instances that have the
highest label very heavily for under-predictions.
Conversely, we do not penalize the meta-classifier for over
predicting.

There has also been recent work on building cost sensitive regression
algorithms~{\citep{zsb11-dss}}, but this is still an active area of research and
beyond the scope of our work.
Instead, we embrace and extend another common technique in regression ---
choosing a fixed threshold and creating a binary classifier.
However, we found that a single threshold was not sufficient for our needs,
and that the approach could be extended to make a series of binary
predictions to find the best cutoffs.
We explain the mechanics of this technique next.

\myparagraph{Cascaded Classification.}
Our approach to prediction relies on a cascade of binary classifiers.
Since classes are ordinal and should be treated as such, a series of
binary predictions can be used to find the minimum cutoff for each
query that also maximizes the overall effectiveness in the final
document reordering stage.
Our approach is similar in spirit to the cascade of classifiers
developed by {\citet{cw+12-aistats}}, and later extended by~{\citet{cx+14-jmlr}}
to minimize the costs of feature evaluation.

\begin{algorithm}
\caption{\sc MultiClassToBinary}
\mylabel{alg:mc2b}

\begin{algorithmic}[1]
\Require A set of queries $Q$, and an
optimal cutoff selected from $c$ choices. 

\Ensure A total of $c-1$ binary training sets $\mathcal{B}$

\For{$i = 1$ {\bf to} $c-1$}
  \For{$q = 0$ {\bf to} $|Q|$}
  \If{{\sc Class}($q) \le c$}
    \State {\sc Class} ($\mathcal{B}_i$[$q$]) $\leftarrow 0$ 
  \Else
    \State {\sc Class}($\mathcal{B}_i$[$q$]) $\leftarrow 1$ 
  \EndIf
  \EndFor
\EndFor
\State {\bf Return} $\mathcal{B}$
\end{algorithmic}
\end{algorithm}

In this work, a random forest classifier~{\citep{b01-ml}} is trained
and used for predictions at each stage of the cascade.
Before building the classifier, training sets can be created from a
multilabeled class set.
The number of binary classifiers required is $c-1$ where $c$ is the
maximum ordinal label.
Labels should be monotonically increasing from $1$ to $c$.
Algorithm~{\ref{alg:mc2b}} shows the approach used to generate a
training set that can be used for iterated binary classifications.

\begin{figure}[t]
\centering
\includegraphics[width=\onecolfigl\textwidth]{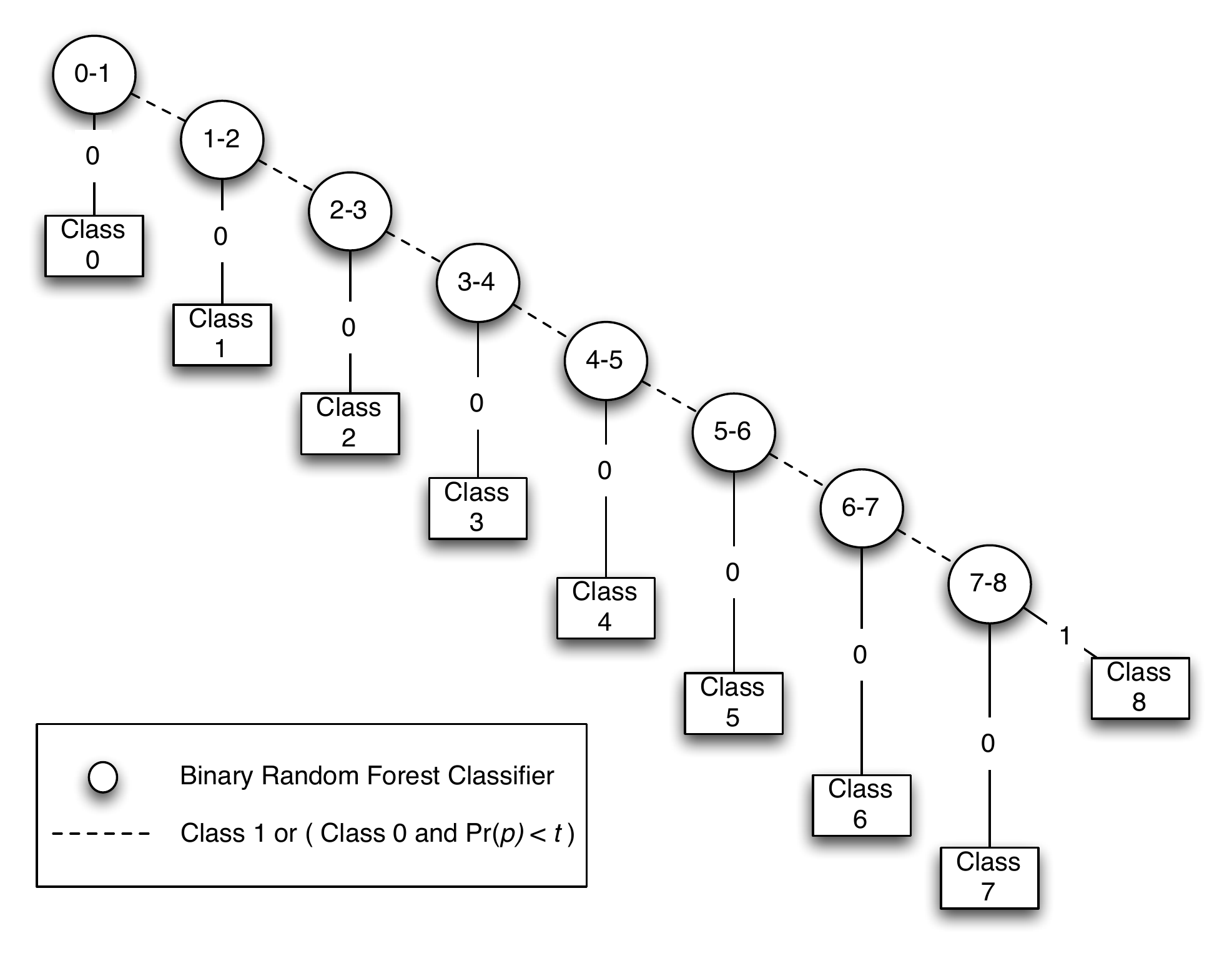}
  \caption{A left-to-right nine class cascade. Each node is a
  binary, random forest classifier. If the classifier predicts
  $0$ with a probability $\var{Pr}(p) > t$, the current node ID is written 
  as the class. Otherwise, the instance is passed to the next
  classifier in the chain.}
  \label{fig:cascade}
\end{figure}

\begin{algorithm}
\caption{\sc LRCascade}
\mylabel{alg:lrcascade}

\begin{algorithmic}[1]
\Require A query $q$, a minimum confidence threshold $t$, and a set of $c-1$
binary classifiers $\mathcal{C}$

\Ensure A cutoff prediction between $1$ and $c$.

\For{$i = 1$ {\bf to} $c-1$}
  \State $p \leftarrow$ {\sc Predict}($C_i$,$q$)
  \If{$p = 0$ {\bf and} $Pr(p) > t$}
    \State {\bf Return} $i$
  \EndIf
\EndFor
\State {\bf Return} $c$
\end{algorithmic}
\end{algorithm}

Once the binary classifiers are constructed, it is relatively simple
to make a prediction for any query $q$.
A feature set can easily be constructed at query parsing time which
is then used by Algorithm~{\ref{alg:lrcascade}} to assign a cutoff
for the query.
A left-to-right cascade serves two important purposes.
First, the model implicitly minimizes the likelihood of a false
positive as assignments are made smallest to largest, and exits
only occur for high probability predictions.
Secondly, each prediction has a small cost.
In a left-to-right cascade, queries with the smallest cutoff incur
the least amount of processing time.
If a larger cutoff is required, the cost of extra predictions is
small relative to the cost of more expensive reordering stages
of large candidate sets later in the scoring process.

Figure~{\ref{fig:cascade}} shows an example of a left-to-right nine 
class cascade of binary classifiers.
Each node in the tree is a binary random forest classifier pre-trained
using one of $\mathcal{B}$ training sets generated using 
Algorithm~{\ref{alg:mc2b}}.
By increasing the cutoff threshold $t$, the percentage of under-predictions
is decreased at the cost of increasing the percentage of over-predictions.
However, some level of over-prediction is always acceptable as this
always results in a gradual increase in overall effectiveness.

\section{Experiments}

\myparagraph{Experimental Configuration.}
For all experiments, $40{,}000$ queries from the 2009 Million Query
Track (MQ2009) were used with a stopped and unpruned
ClueWeb 2009 category B index (CW09B).
The {\tt{uogTRMQdph40}} system is used as the gold standard, as it
represents one of the top-scoring systems (when measured over the
small subset of the queries that were evaluated) that returned runs
for all $40{,}000$ MQ2009 queries.
Specifically, this is the highest scoring system that submitted
results for all of the queries in MQ2009, making it the best choice
as the gold standard in our work.

To generate the bag-of-words candidate run, a BM25 implementation
using the same formulation and parameterization as described in
Section~{\ref{sec-approach}} was ran for all $40{,}000$ MQ2009 queries.
The stopword list and Krovetz stemmer were derived directly from the
{\indri}\footnote{\url{http://www.lemurproject.org/indri.php}} search
engine.
A total of $50{,}22{,}0423$ documents were indexed from the
CW09B collection, and all queries were ran to a depth of
$10{,}000$.

For classification, the $k$ and $\rho$ values were computed at nine 
different positions. For $k$, the values were $20$, $50$, $100$,
$200$, $500$, $1{,}000$, $2{,}000$, $5{,}000$, and $10{,}000$.
For $\rho$, the values were $100$k, $200$k, $500$k, $1$m, $2$m,
$5$m, $10$m, $20$m, and $50$m.
For each bucket, three different {\MED} variants were computed:
$\med{RBP}$, $\med{ERR}$, and $\med{DCG}$.
Cross validation was performed by partitioning all of the
queries into $10$ folds using the {\tt StratifiedRemoveFolds} filter
in {\method{Weka-3.7.13}}.
Then, $10$ runs of each classification approach were ran using 
$9$ folds for training, and the current fold for testing to generate
a prediction for each topic in the collection.
Note that before generating the final folds,
we removed all queries for which we had any judgments ($687$ topics) at all,
for further validation purposes.

\begin{figure*}[t]
\centering
\includegraphics[width=0.49\textwidth]{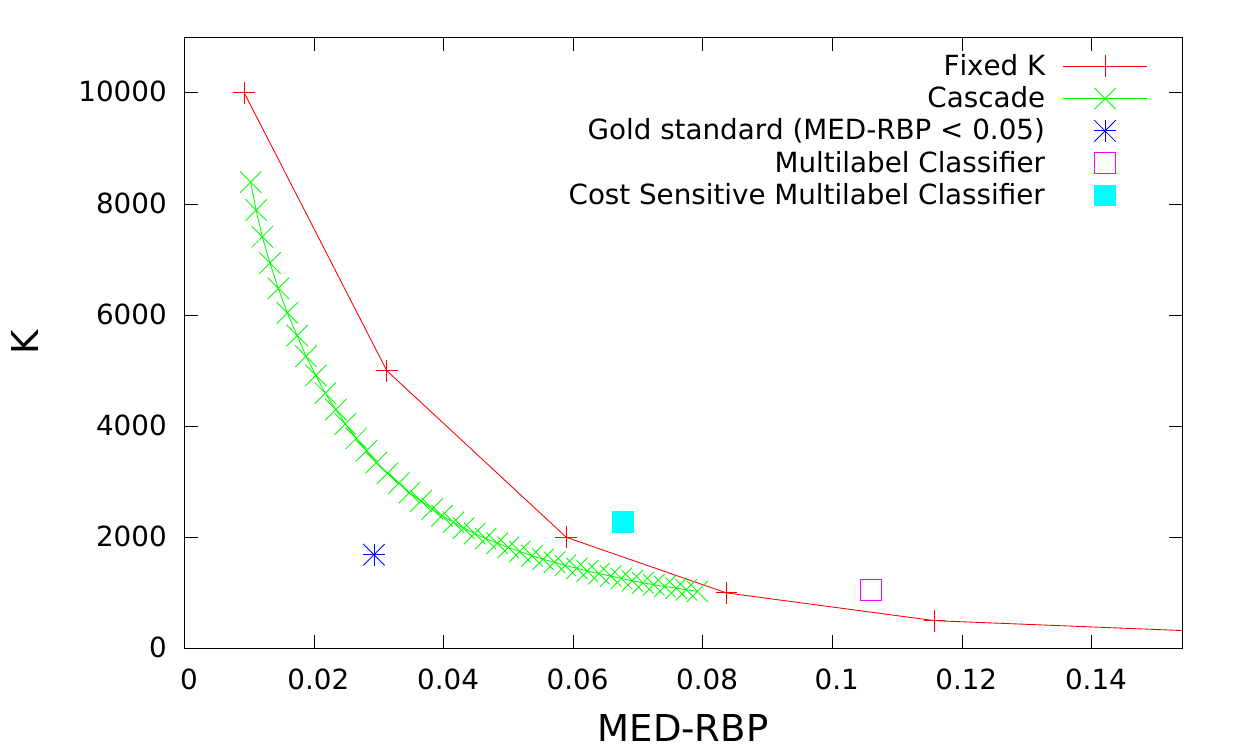}
\includegraphics[width=0.49\textwidth]{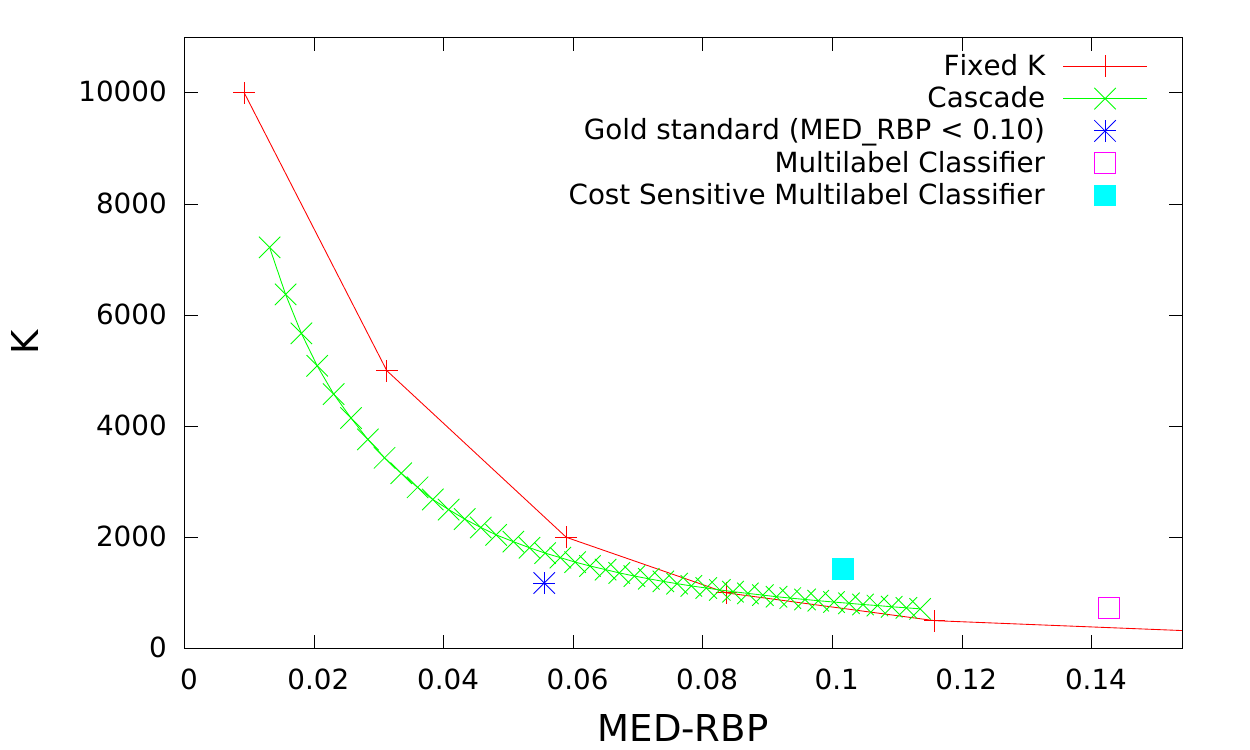}
  \caption{$\med{RBP}$ versus $k$ when using a training threshold
  cutoff of $\med{RBP}\le 0.05$ (left panel) and $\med{RBP}\le 0.10$
  (right panel) for the MQ2009 queries on ClueWeb09B. The blue
  star represents the best possible result achievable with a ``perfect''
  classifier, the green line is the result when using our LR Binary
  Cascade Model, and the red line represents the tradeoff horizon
  based on using a fixed $k$ for all queries.
  }
  \label{fig:krbp}
\end{figure*}

\myparagraph{Dynamic Selection of $k$.}
Our first set of experiments were designed to test the hypothesis that
a best $k$ value can be determined on a query by query basis which
{\em minimizes} effectiveness loss, and {\em maximizes} efficiency.
In other words, finding the smallest acceptable $k$ for a targeted
{\MED} value can minimize the amount of work later stage rerankers
must do, and also minimizes the cost of using a safe-to-$k$ candidate
generation algorithm such as {\sc Wand}.
In order to prove this hypothesis, we created several different
datasets to train our predictor.
We experimented with $\med{RBP}$ using the cutoffs $0.02$, $0.05$,
$0.10$, $0.15$, $0.20$, $0.25$, $0.30$, and $0.50$.
We also experimented with $\med{DCG}$ using the cutoffs 
$0.2$, $0.3$, $0.5$, $0.7$, $1.00$, $1.20$, and $1.50$.
Similar experiments were performed with $\med{ERR}$ 
using the cutoffs $0.05$, $0.10$, $0.15$, $0.20$, $0.25$, $0.30$, and $0.50$.

Figure~{\ref{fig:krbp}} shows the tradeoff achievable between
the candidate set size $k$ and $\med{RBP}$.
The left pane is a summary of results when using a target of 
$\med{RBP} \le 0.05$, and the right pane summarizes method
performance for $\med{RBP} \le 0.10$.
In both graphs, the red line represents the tradeoff horizon in
efficiency and effectiveness when using a fixed cutoff for all
queries as is generally done in current system configurations.
The blue star represents the gold standard result that would be
achievable with a ``perfect'' classifier.
The two squares represent the result achievable when using standard
machine learning approaches such as a Bayesian Model Combination, 
boosted, multilabel random forest classifier~{\citep{mc+11-ijcnn}}
\footnote{\url{http://uaf46365.ddns.uark.edu/waffles/}}, or a 
Cost Sensitive Classifier such as \method{MetaCost}~{\citep{d99-kdd}}
\footnote{\url{http://www.cs.waikato.ac.nz/ml/weka/}}. 
When compared to the fixed baseline, we see that a traditional
approach to classification does not provide any real benefit.

In contrast, the {\method{LRCascade}} approach (green line) shows
clear improvements over both multi-label and fixed cutoff approaches.
The lower the choice of {\MED}, the less likely there is any loss in
effectiveness.
Our experiments suggest that targeting low {\MED} values are
likely to reap the most rewards.
This is, the process of minimizing effectiveness loss 
greatly benefits from a variable cutoff approach.

\begin{table*}[t]
\centering
\begin{small}
\begin{tabular}{llccccccccc}
\toprule
\multirow{4}{*}{Method} &&
  \multicolumn{4}{c}{Interpolated ${\med{RBP}}$} &&
    \multicolumn{4}{c}{Interpolated $k$} \\
  \cmidrule{3-6}
    \cmidrule{8-11}
  && Predicted
    & Predicted
      & Fixed
        & Difference in
          &
            & Predicted
              & Predicted
                & Fixed
                  & Difference in \\
  && ${\med{RBP}}$ 
    & $k$ 
      & $k$
        & $k$
          &
            & $k$ 
              & ${\med{RBP}}$ 
                & ${\med{RBP}}$ 
                  & ${\med{RBP}}$ \\
\midrule
\method{Oracle}
  &
  & $0.029$ 
    & $1{,}688$ 
      & $5{,}459$
        & +$223\%$
          &
            & $1{,}688$ 
              & $0.029$ 
                & $0.067$ 
                  & +$128\%$ \\
\method{MultiLabel}
  &
  & $0.106$ 
    & $1{,}053$ 
      & $\C653$
        & --$\D38\%$
          &
            & $1{,}053$ 
              & $0.106$ 
                & $0.082$ 
                  & --$\D22\%$ \\
\method{MetaCost}
  &
  & $0.068$ 
    & $2{,}277$ 
      & $1{,}644$
        & --$\D28\%$
          &
            & $2{,}277$ 
              & $0.068$ 
                & $0.056$ 
                  & --$\D16\%$ \\
\method{LRCascade}, $t=0.75$
  &
  & $0.045$ 
    & $2{,}071$ 
      & $3{,}535$
        & +$\D71\%$
          &
            & $2{,}071$ 
              & $0.045$ 
                & $0.058$ 
                  & +$\D30\%$ \\
\method{LRCascade}, $t=0.80$
  &
  & $0.036$ 
    & $2{,}656$ 
      & $4{,}432$
        & +$\D67\%$
          &
            & $2{,}656$ 
              & $0.036$ 
                & $0.053$ 
                  & +$\D45\%$ \\
\method{LRCascade}, $t=0.85$
  &
  & $0.028$ 
    & $3{,}561$ 
      & $5{,}715$
        & +$\D61\%$
          &
            & $3{,}561$ 
              & $0.028$ 
                & $0.044$ 
                  & +$\D59\%$ \\
\bottomrule
\end{tabular}
\caption{Interpolated $k$ and ${\med{RBP}}$ when training at
${\med{RBP}} \le 0.05$. The relative gain or loss for $k$ and
${\med{RBP}}$ are shown when compared to using a fixed cutoff
for all queries. The Oracle method represents the best
possible result given a perfect classifier.
\label{tbl:kgains}
}
\end{small}
\end{table*}

Table~{\ref{tbl:kgains}} shows the breakdown for using a fixed $k$
and predicted $k$ interpolation when using a training set
targeted at $\med{RBP} \le 0.05$.
Columns $2$--$5$ show the relative gain in terms of $k$, and columns
$6$--$9$ show the relative gain in terms of $\med{RBP}$.
The first row shows the gold standard {\method{Oracle}} result, which
represents the best result that is achievable using this
parameter -- metric -- target threshold combination.
Changing any one of these three constraints will change the gain (or
loss) possible.
In other words, just computing the {\method{Oracle}} result is in
itself interesting, as it provides a {\it bound} on how much
benefit the three constraint combination could provide.

The interpretation of the data in columns $2$--$5$ is as
follows:\ given a particular setting, how far {\it below} the
interpolated fixed $k$ curve (red) are we? That is, if we accept a
particular level of $\med{RBP}$ effectiveness, how much efficiency can we gain
over simply just adopting a fixed $k$ cutoff for all queries 
(specifically, the $k$ cutoff that would achieve the same level of $\med{RBP}$)?
The interpretation of the data in columns $6$--$9$ is as follows:\ given a
particular setting, how far {\it left} of the interpolated fixed $k$
curve (red) are we? That is, how much more effective (in terms of
$\med{RBP}$) can we make our results over simply setting a fixed
$k$? In designing actual search architectures, the first
interpretation is more intuitive, since we want to optimize efficiency
without sacrificing effectiveness, but the alternative perspective is
interesting as well in quantifying the benefits of our technique.

We see that both {\method{MultiLabel}} and {\method{MetaCost}} are
marginally worse than a fixed cutoff, with {\method{MetaCost}} being
the slightly better choice.
The {\method{LRCascade}} method is the clear winner across a wide
range of $t$.
The exact value of $t$ can be set depending on which direction a user
wishes to bias the tradeoff.
Choosing a lower $t$ decreases the average $k$, while increasing the
average $\med{RBP}$.
Choosing a higher $t$ biases the tradeoff in the effectiveness
direction.

\begin{figure*}[t]
\centering
\includegraphics[width=0.49\textwidth]{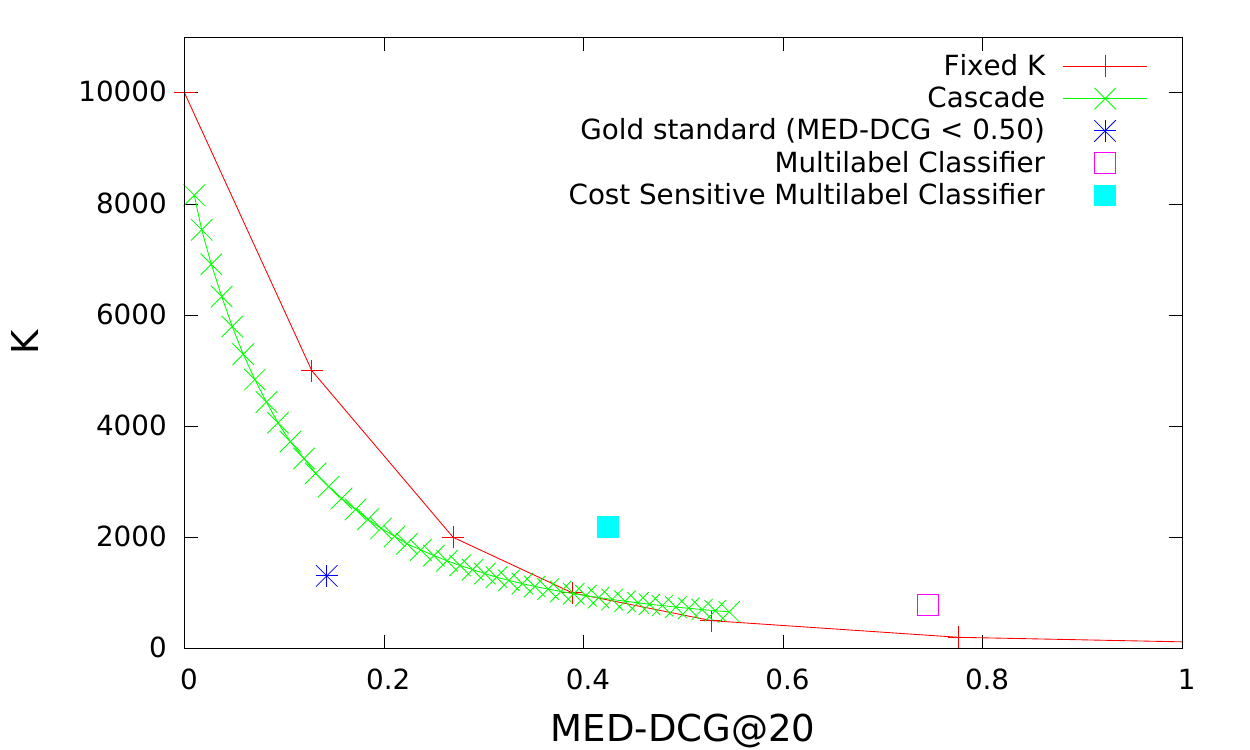}
\includegraphics[width=0.49\textwidth]{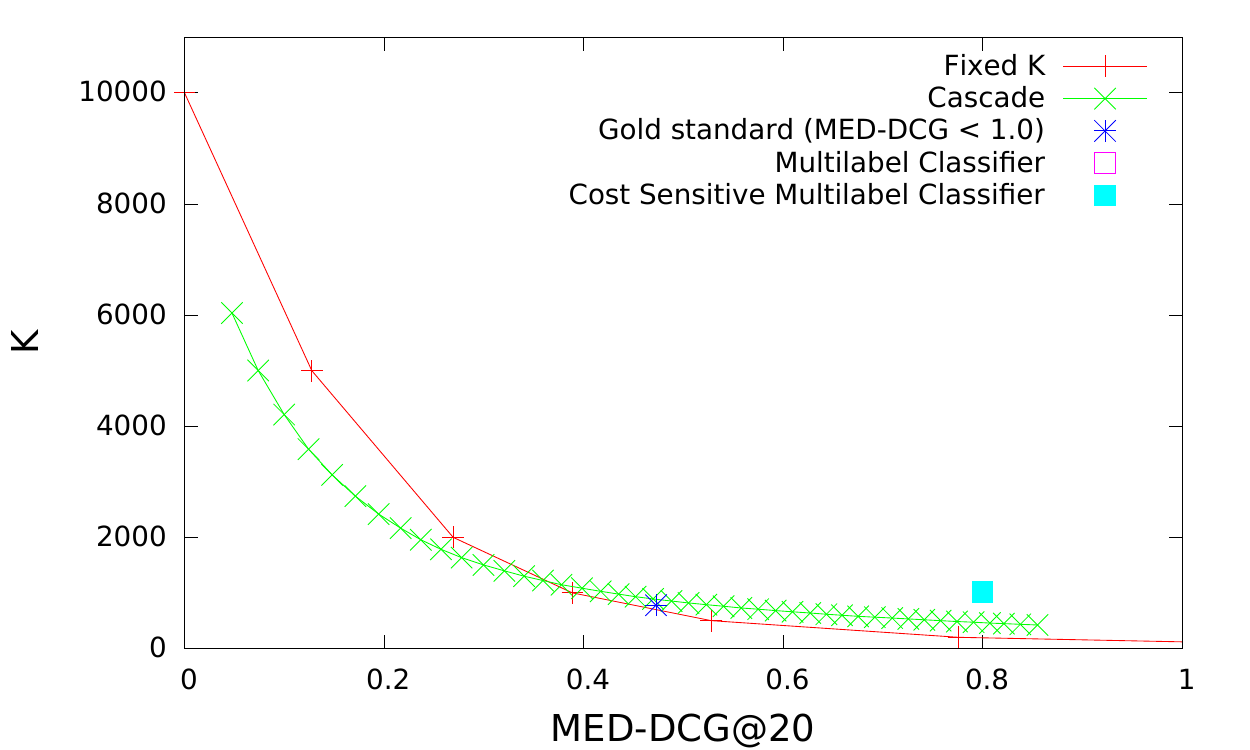}
  \caption{$\med{DCG}$ versus $k$ when using a training threshold
  cutoff of $\med{DCG}\le 0.50$ (left panel) and $\med{DCG}\le 1.0$
  (right panel) for the MQ2009 queries on ClueWeb09B. The blue
  star represents the best possible result achievable with a ``perfect''
  classifier, the green line is the result when using our LR Binary
  Cascade Model, and the red line represents the tradeoff horizon
  based on using a fixed $k$ for all queries.
  }
  \label{fig:kdcg}
\end{figure*}

Figure~{\ref{fig:kdcg}} shows the same experiment when using 
$\med{DCG} \le 0.50$ and $\med{DCG} \le 1.00$.
Changing the underlying evaluation metric does not change
the general trends for all methods tested.
Multilabel classifiers do not outperform fixed cutoffs, while
the {\method{LRCascade}} is the superior tradeoff.
We also ran a similar set of experiments using $\med{ERR}$
and achieved similar results and trends.

\begin{figure*}[t]
\centering
\includegraphics[width=0.49\textwidth]{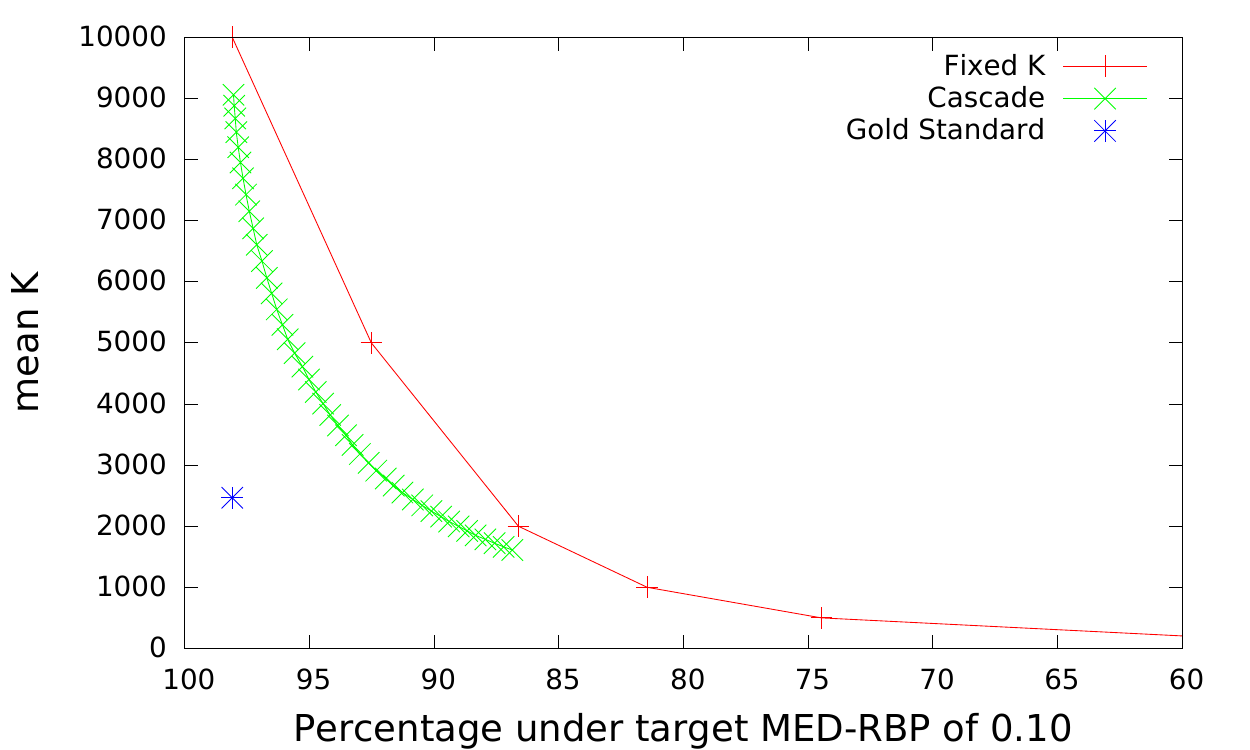}
\includegraphics[width=0.49\textwidth]{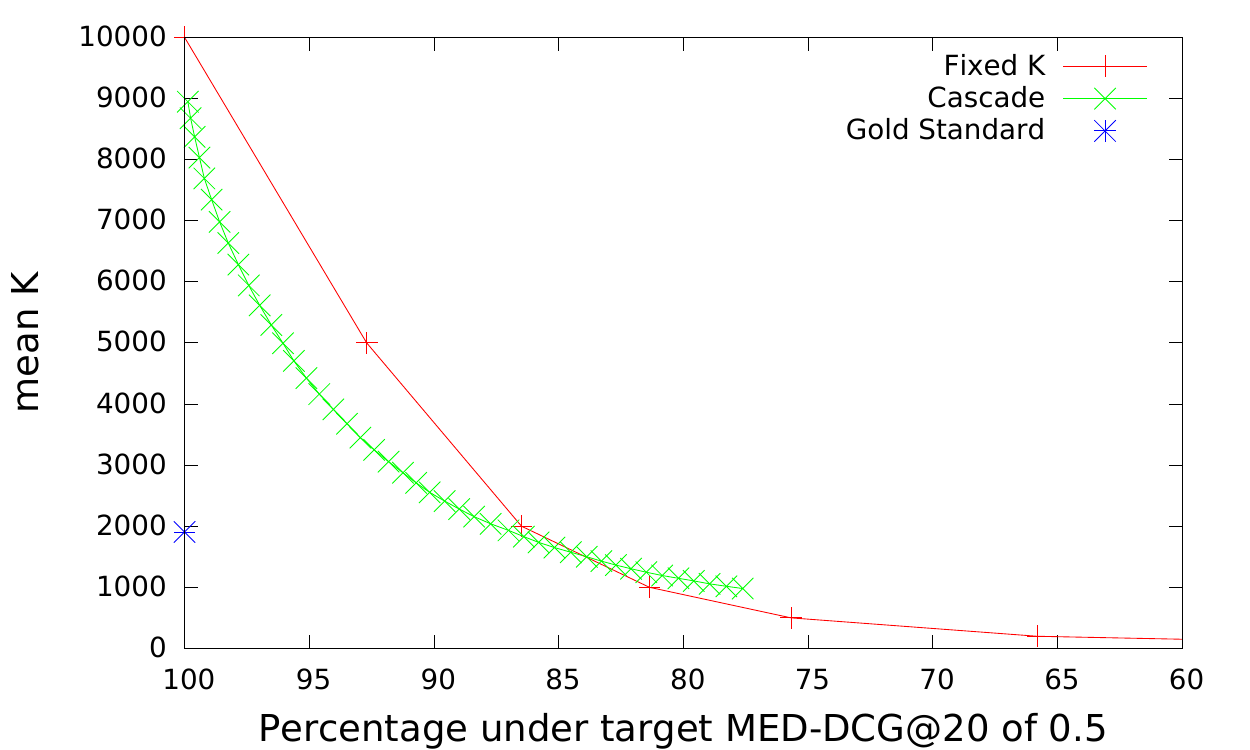}
  \caption{Average $k$ versus the percentage of queries achieving
  a {\MED} score less than a training target.
  The left panel shows the comparison of the {\method{LRCascade}}
  (green line) and a fixed cutoff (red line) for $\med{RBP} \le 0.10$.
  The right panel shows the same comparison when using 
  $\med{DCG} \le 0.50$.
  }
  \label{fig:under}
\end{figure*}

Figure~{\ref{fig:under}} shows the percentage of queries which
obtain a bound of $\med{RBP} \le 0.10$ or $\med{DCG} \le 0.50$.
We can see that the {\method{LRCascade}} approach is clearly
predicting cutoffs which have a lower mean $k$, and a higher
percentage of queries under the target {\MED}, translating
into better effectiveness.
You may notice that even the gold standard does not achieve 100\%
under $\med{RBP}$.
For a proportion of the topics our first stage returns less than
the target $K$ documents due to a lack of documents containing any 
of the query terms.
Since $\med{RBP}$, like RBP, is conceptually evaluated to infinite
depth, this deficiency is reflected by positive scores,
some of which fall above the targeted value.
On the other hand, $\med{DCG}$ is evaluated to fixed depth
(depth 20 in this case) and the gold standard achieves 100\%.

\begin{table*}[t]
\centering
\begin{small}
\begin{tabular}{llccccccccc}
\toprule
\multirow{4}{*}{Method} &&
  \multicolumn{4}{c}{Interpolated ${\med{ERR}}$} &&
    \multicolumn{4}{c}{Interpolated $k$} \\
  \cmidrule{3-6}
    \cmidrule{8-11}
  && Predicted
    & Predicted
      & Fixed
        & Difference in
          &
            & Predicted
              & Predicted
                & Fixed
                  & Difference in \\
  && ${\med{ERR}}$ 
    & $k$ 
      & $k$
        & $k$
          &
            & $k$ 
              & ${\med{ERR}}$ 
                & ${\med{ERR}}$ 
                  & ${\med{ERR}}$ \\
\midrule
\method{Oracle}
  &
  & $0.017$ 
    & $1{,}752$ 
      & $7{,}344$
        & +$319\%$
          &
            & $1{,}752$ 
              & $0.017$ 
                & $0.067$ 
                  & +$292\%$ \\
\method{MultiLabel}
  &
  & $0.117$ 
    & $1{,}159$ 
      & $\C452$
        & --$\D61\%$
          &
            & $1{,}159$ 
              & $0.117$ 
                & $0.082$ 
                  & --$\D30\%$ \\
\method{MetaCost}
  &
  & $0.060$ 
    & $3{,}222$ 
      & $2{,}214$
        & --$\D31\%$
          &
            & $3{,}222$ 
              & $0.060$ 
                & $0.050$ 
                  & --$\D16\%$ \\
\method{LRCascade}, $t=0.75$
  &
  & $0.047$ 
    & $2{,}206$ 
      & $3{,}465$
        & +$\D57\%$
          &
            & $2{,}206$ 
              & $0.047$ 
                & $0.060$ 
                  & +$\D26\%$ \\
\method{LRCascade}, $t=0.80$
  &
  & $0.035$ 
    & $3{,}013$ 
      & $4{,}705$
        & +$\D56\%$
          &
            & $3{,}013$ 
              & $0.035$ 
                & $0.052$ 
                  & +$\D47\%$ \\
\method{LRCascade}, $t=0.85$
  &
  & $0.024$ 
    & $4{,}191$ 
      & $6{,}351$
        & +$\D52\%$
          &
            & $4{,}191$ 
              & $0.024$ 
                & $0.040$ 
                  & +$\D70\%$ \\
\bottomrule
\end{tabular}
\caption{Interpolated $k$ and ${\med{ERR}}$ when training at
${\med{ERR}} \le 0.05$. The relative gain or loss for $k$ and
${\med{ERR}}$ are shown when compared to using a fixed cutoff
for all queries. The Oracle method represents the best
possible result given a perfect classifier.
\label{tbl:kerrgains}
}
\end{small}
\end{table*}

Finally, Table~{\ref{tbl:kerrgains}} shows the breakdown for several
fixed $k$ and predicted $k$ interpolations when building the training
set to target $\med{ERR} \le 0.05$.
The trends remain consistent as when using $\med{RBP}$ or
$\med{DCG}$.
The most interesting aspect of this table is to note the subtle
difference in potential improvements possible for the gold standard
{\method{Oracle}} result.
Potential gains are +$319\%$ and +$292$ respectively.
This is a little better than the {\method{Oracle}} result for
$\med{RBP}$ shown in Table~{\ref{tbl:kgains}}, or that of
$\med{DCG} \le 0.50$ which shows potential gains of
+$259\%$ for $k$ and +$147$ for {\MED}.
Potential gains are sensitive to both training cutoff
and metric, which should come as no surprise to the
reader.

\begin{figure*}[t]
\centering
\includegraphics[width=0.48\textwidth]{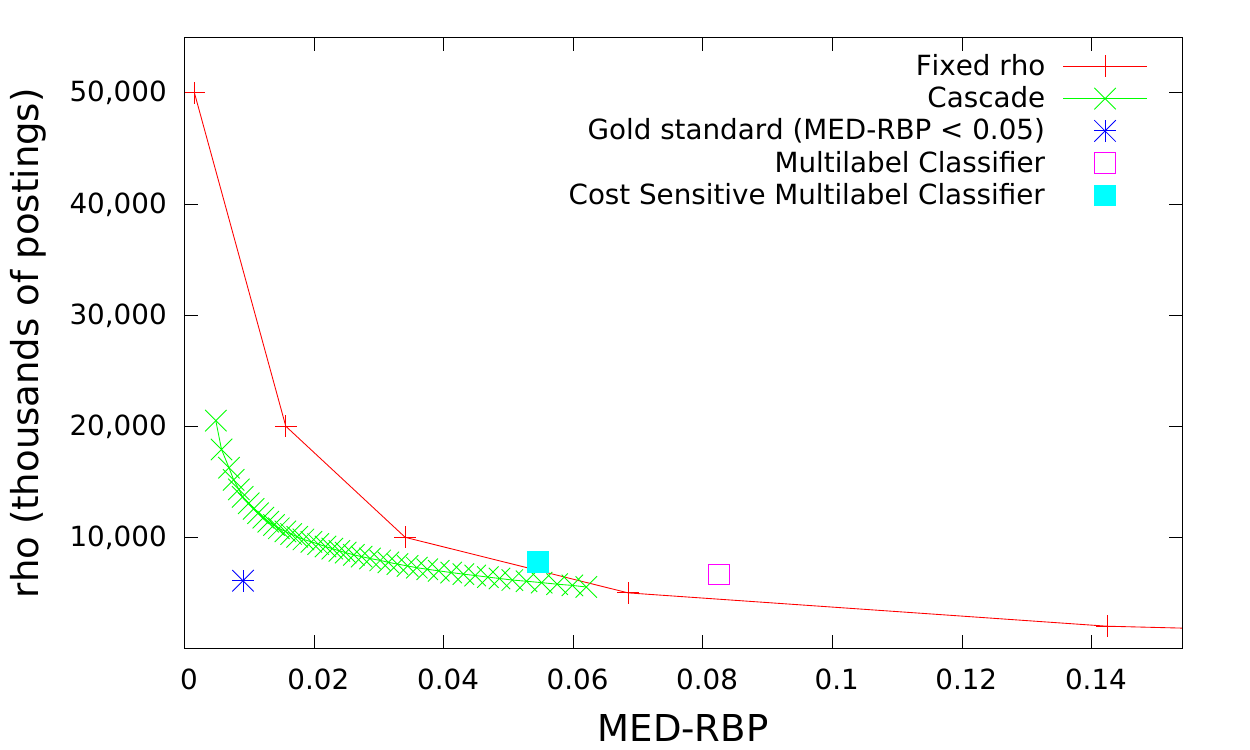}
\includegraphics[width=0.48\textwidth]{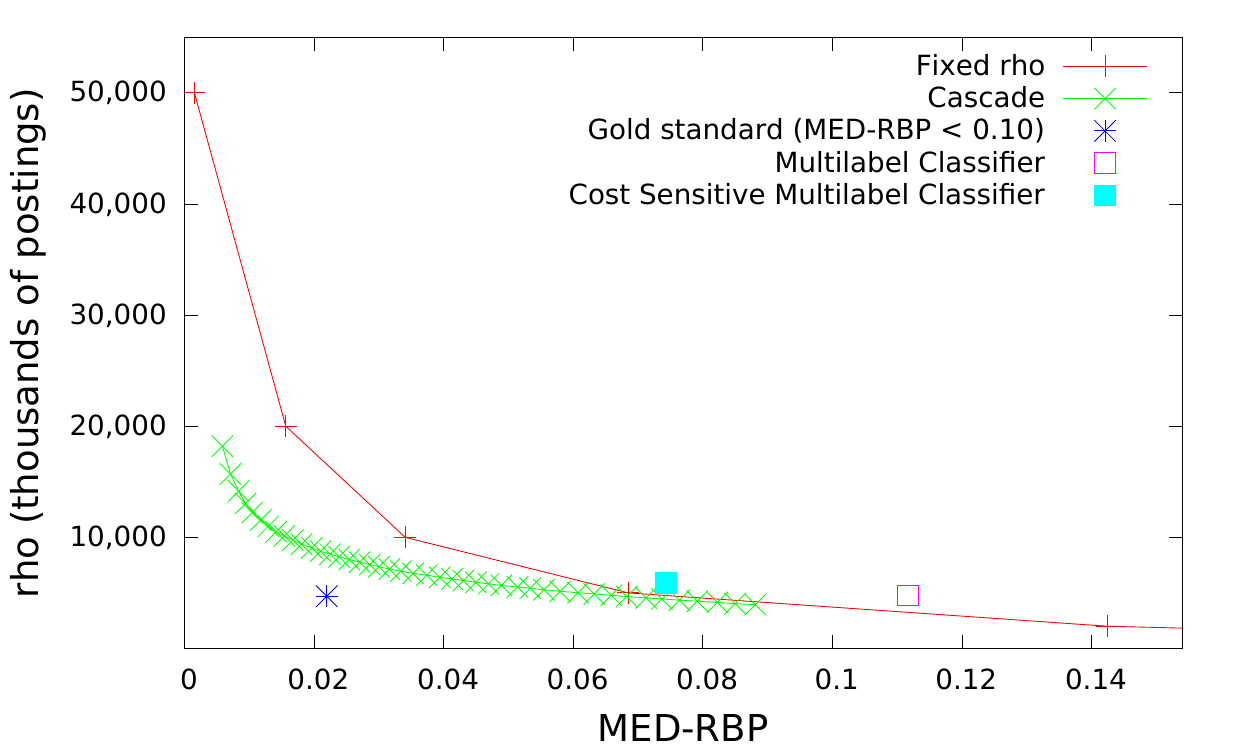}
  \caption{$\med{RBP}$ versus $\rho$ when using a training threshold
  cutoff of $\med{RBP}\le 0.05$ (left panel) and $\med{RBP}\le 0.10$
  (right panel) for the MQ2009 queries on ClueWeb09B. The blue
  star represents the best possible result achievable with a ``perfect''
  classifier, the green line is the result when using our LR Binary
  Cascade Model, and the red line represents the tradeoff horizon
  based on using a fixed $k$ for all queries.
  }
  \label{fig:krho}
\end{figure*}

\myparagraph{Dynamic Selection of $\rho$.}
We now turn our attention to the parameter $\rho$.
The $\rho$ parameter controls the number of postings scored
using a score-at-a-time approximation algorithm. 
Quite simply, it is a parameter that can be used to tune
an efficiency-effectiveness tradeoff.
Our methodology is identical to the approach taken for $k$.
The only difference is the initial generation of the cutoff
data that is used for training the classifier.

Figure~{\ref{fig:krho}} shows the effect of $\med{RBP}$ for
the training cutoffs $0.05$ and $0.10$.
When comparing both graphs we can see that training with a
smaller cutoff provides a clear advantage.
The gold standard point provides an ambitious goal.
Once again the {\method{LRCascade}} approach is clearly
better than the fixed cutoff in both dimensions,
{\MED} and $\rho$.

\begin{table*}[t]
\centering
\begin{small}
\begin{tabular}{llccccccccc}
\toprule
\multirow{4}{*}{Method} &&
  \multicolumn{4}{c}{Interpolated ${\med{RBP}}$} &&
    \multicolumn{4}{c}{Interpolated $\rho$} \\
  \cmidrule{3-6}
    \cmidrule{8-11}
  && Predicted
    & Predicted
      & Fixed
        & Difference in
          &
            & Predicted
              & Predicted
                & Fixed
                  & Difference in \\
  && ${\med{RBP}}$ 
    & $\rho$ 
      & $\rho$
        & $\rho$
          &
            & $\rho$ 
              & ${\med{RBP}}$ 
                & ${\med{RBP}}$ 
                  & ${\med{RBP}}$ \\
\midrule
\method{Oracle}
  &
  & $0.009$ 
    & $\D6{,}094$ 
      & $34{,}085$
        & +$459\%$
          &
            & $\D6{,}094$ 
              & $0.009$ 
                & $0.061$ 
                  & +$575\%$ \\
\method{MultiLabel}
  &
  & $0.082$ 
    & $\D6{,}656$ 
      & $\D4{,}434$
        & --$\D33\%$
          &
            & $\D6{,}656$ 
              & $0.082$ 
                & $0.057$ 
                  & --$\D31\%$ \\
\method{MetaCost}
  &
  & $0.054$ 
    & $\D7{,}787$ 
      & $\D7{,}032$
        & --$\D10\%$
          &
            & $\D7{,}787$ 
              & $0.054$ 
                & $0.049$ 
                  & --$\D\D9\%$ \\
\method{LRCascade}, $t=0.75$
  &
  & $0.026$ 
    & $\D8{,}430$ 
      & $14{,}350$
        & +$\D70\%$
          &
            & $\D8{,}430$ 
              & $0.026$ 
                & $0.045$ 
                  & +$\D72\%$ \\
\method{LRCascade}, $t=0.80$
  &
  & $0.021$ 
    & $\D9{,}388$ 
      & $17{,}318$
        & +$\D84\%$
          &
            & $\D9{,}388$ 
              & $0.021$ 
                & $0.038$ 
                  & +$\D86\%$ \\
\method{LRCascade}, $t=0.85$
  &
  & $0.016$ 
    & $10{,}532$ 
      & $20{,}239$
        & +$\D92\%$
          &
            & $10{,}532$ 
              & $0.016$ 
                & $0.033$ 
                  & +$113\%$ \\
\bottomrule
\end{tabular}
\caption{Interpolated $\rho$ and ${\med{RBP}}$ when training at
${\med{RBP}} \le 0.05$. The relative gain or loss for $\rho$ and
${\med{RBP}}$ are shown when compared to using a fixed cutoff
for all queries. The Oracle method represents the best
possible result given a perfect classifier. Note that 
$\rho$ values are in thousands of postings scored.
\label{tbl:rhogains}
}
\end{small}
\end{table*}

Table~{\ref{tbl:rhogains}} shows the $\med{RBP}$ and $\rho$
interpolation results when training the classifier with a cutoff of
$\med{RBP} \le 0.05$.
Again the results are consistent with the general trends observed in
Table~{\ref{tbl:kgains}}.
One key difference is that the potential gain in the tradeoff
achievable by the {\method{Oracle}} method relative to using a fixed
cutoff is $2\times$ in terms of efficiency, and $4\times$ in
effectiveness.
This translates directly into higher relative gains achievable using
our {\method{LRCascade}} approach.

There are several interesting insights that can be gleaned from this
experiment.
Firstly, our classification approach is generalizable.
We use exactly the same feature sets, algorithms, and prediction
methodology in both experiments.
The only key difference is in the construction of the training data.

Perhaps further improvements could be realized by tuning
a number of different configuration options such as the number of class
cutoffs, using variable cutoff thresholds $t$ at different nodes in
the cascade, changing the classifier algorithms (perhaps even
using different classifiers) at different nodes in the
cascade, or even developing an entirely new approach to cascaded
regression / classification.
Initial efforts towards variable cutoff thresholds show promising results.
The gains achievable are independent to all of these decisions.
In fact the precise gain can be computed based on the creation
of the {\method{Oracle}} run before investing any time and effort
into engineering a feature set and classification scheme.

\begin{table}[t]
\centering
\begin{small}
\begin{tabular}{lccr}
\toprule
Method & NDCG@10 & ERR &  \multicolumn{1}{c}{k} \\
\midrule
\method{Oracle} 
& 0.356 & 0.434 & 2,386  \\
\method{LRCascade}, $t=0.75$ 
& 0.359 & 0.435 & 3,422  \\
\method{LRCascade}, $t=0.80$ 
& 0.359 & 0.435 & 4,062  \\
\method{LRCascade}, $t=0.85$ 
& 0.358 & 0.435 & 5,130  \\
\method{Fixed, $k = 10,000$}
& 0.358 & 0.434 & 10,000\\
\bottomrule
\end{tabular}
\caption{
Measured performance over 50~held out TREC 2009 Web Track adhoc queries.
\label{tbl:effect}
}
\end{small}
\end{table}

\myparagraph{Validation}
As a final step, we confirmed our past experience
(as illustrated by Figure~\ref{fig:med})
that low $\med{RBP}$ values produce minimal loss in measured effectiveness.
For this purpose we employed the 50~queries of the TREC 2009 Web Track
adhoc task, which were held out from the training and test sets of
other experiments reported in this section.
These 50~queries were pooled to depth 12 for judging,
and so should at least be suitable for computing early-precision
effectiveness measures, including NDCG@10 and ERR.

Table~\ref{tbl:effect} shows the results.
Over these queries,
our cascade classifier produces no measurable loss in
effectiveness when compared to a fixed $k$ of $10{,}000$.
In fact, the classifer achieves a tiny (but not significant) gain
in effectiveness in the third decimal place of some measures,
reflecting a change of one or two documents across this small query set.
On the other hand, there are substantial reductions in average $k$,
reflecting expected efficiency improvements.

\section{Conclusion}

In this work, we have presented a new query specific approach to
dynamically predict the best parameter cutoffs that maximises both 
efficiency and effectiveness. 
To achieve this, we use Maximized Effectiveness Difference (\MED)
\citep{tc15,ccm16irj} as the basis for evaluating the quality
of a candidate set relative to a more expensive gold standard
reranking step.
By extending this methodology, we are able to create a large test
corpora, and train a remarkably robust classifier which requires
{\em no} relevance judgements.
Our approach to binary cascaded classification is able to achieve up
to a $50\%$ improvement in average $k$.
For $\rho$, we achieve up to an even greater relative improvement in
average number of postings scored.
Our approach can easily be generalized to effectively tune a wide
variety of other parameters dynamically in multi-stage retrieval
systems, and can be used to reliable estimate potential gains
achievable with any parameter --- metric --- target threshold
combination.

\myparagraph{Acknowledgments.}
This work was partially supported by
the Natural Sciences and Engineering Research Council of Canada (NSERC),
and by
the Australian Research Council's
{\emph{Discovery Projects}} Scheme (DP140103256).
Shane Culpepper is the recipient of an Australian Research Council
DECRA Research Fellowship (DE140100275).

\balance
\InputIfFileExists{p.bbl}


\end{document}